\documentclass[pra, reprint, superscriptaddress, amsmath, amssymb, aps, floatfix]{revtex4-2}
\usepackage{orcidlink}
\usepackage{tabularray} 

\usepackage{physics}
\usepackage{siunitx}

\usepackage{siunitx}
\usepackage{graphicx}
\usepackage{dcolumn}
\usepackage{rotating}          
\usepackage{bm}

\usepackage[utf8]{inputenc}
\usepackage[T1]{fontenc}

\usepackage{booktabs}
\usepackage{multirow}
\usepackage{diagbox}

\usepackage[normalem]{ulem}

\def\Rb87{^{87}\mathrm{Rb}}     
\def\Na23{^{23}\mathrm{Na}}     
\def\K39{^{39}\mathrm{K}}                    		    
\def\kB{k_{\rm B}}                            			
\def\FVU{\mathrm{FVU}}

\def\ex{\mathbf{e}_x}  
\def\ey{\mathbf{e}_y}  
\def\ez{\mathbf{e}_z}

\def\SNR{{\rm SNR}}

\begin{document}
\title{Nondestructive characterization of laser-cooled atoms using machine learning}

\author{G.~De~Sousa\ \orcidlink{0000-0002-8529-5439}}
\thanks{These authors contributed equally to this work.}
\affiliation{Department of Physics$,$ University of Maryland$,$ College Park$,$ MD 20742$,$ USA}
\affiliation{National Institute of Standards and Technology, Gaithersburg, Maryland, 20899, USA}
\affiliation{Instituto de Física de São Carlos$,$ Universidade de São Paulo$,$ 13560-970$,$ São Carlos$,$ SP$,$ Brasil}

\author{M.~Doris \orcidlink{0009-0000-5844-8148}}
\thanks{These authors contributed equally to this work.}
\affiliation{Joint Quantum Institute, National Institute of Standards and Technology and University of Maryland, Gaithersburg, Maryland 20899, USA}
\affiliation{National Institute of Standards and Technology, Gaithersburg, Maryland, 20899, USA}

\author{D.~D'Amato\ \orcidlink{0009-0002-7191-9228}}
\thanks{These authors contributed equally to this work.}
\affiliation{Joint Quantum Institute, National Institute of Standards and Technology and University of Maryland, Gaithersburg, Maryland 20899, USA}
\affiliation{National Institute of Standards and Technology, Gaithersburg, Maryland, 20899, USA}

\author{B.~Egleston\ \orcidlink{0009-0004-7516-7689}}
\affiliation{Joint Quantum Institute, National Institute of Standards and Technology and University of Maryland, Gaithersburg, Maryland 20899, USA}
\affiliation{National Institute of Standards and Technology, Gaithersburg, Maryland, 20899, USA}

\author{J.~P.~Zwolak\ \orcidlink{0000-0002-2286-3208}}
\email{jpzwolak@nist.gov}
\affiliation{National Institute of Standards and Technology, Gaithersburg, Maryland, 20899, USA}

\author{I.~B.~Spielman\ \orcidlink{0000-0003-1421-8652}}
\email{ian.spielman@nist.gov}
\affiliation{National Institute of Standards and Technology, Gaithersburg, Maryland, 20899, USA}
\affiliation{Joint Quantum Institute, National Institute of Standards and Technology and University of Maryland, Gaithersburg, Maryland 20899, USA}

\date{\today}

\begin{abstract}
We develop machine learning techniques for estimating physical properties of laser-cooled potassium-39 atoms in a magneto-optical trap using only the scattered light---i.e., fluorescence---that is intrinsic to the cooling process.
In-situ snap-shot images of fluorescing atomic ensembles directly reveal the spatial structure of these millimeter-scale objects but contain no obvious information regarding internal properties such as the temperature.
We first assembled and labeled a balanced dataset sampling $8\times10^3$ different experimental parameters that includes examples with: large and dense atomic ensembles, a complete absence of atoms, and everything in between.
We describe a range of models trained to predict atom number and temperature solely from fluorescence images.
These run the gamut from a poorly performing linear regression model based only on integrated fluorescence to deep neural networks that give number and temperature with fractional uncertainties of $0.1$ and $0.2$ respectively.
\end{abstract}

\maketitle

\section{Introduction}

The past decade has witnessed a rapid adoption of machine learning (ML) techniques in the applied and fundamental physical sciences~\cite{Carleo2019}.
These approaches have been used for everything 
from stabilizing nuclear fusion reactors~\cite{Degrave2022} and designing and controlling quantum devices~\cite{Zwolak2023}, to imaging black-hole event horizons~\cite{Medeiros2023}, discovering new materials~\cite{Merchant2023}, and searching for physics beyond the standard model of particle physics~\cite{Karagiorgi2022}.
A key use case is the identification of ``hidden'' information: for example, in many-body quantum systems, topological order is ``hidden'' because its signatures are highly non-local~\cite{Hasan2010}.
Nevertheless, ML tools have demonstrated the ability to identify these phases with both simulated~\cite{Carrasquilla2017} and experimental data~\cite{Rem2019}.
Our focus is analogous: estimating internal properties of laser-cooled atoms from purely non-destructive fluorescence images, where only scattered light intrinsic to the cooling process is observed.
While such fluorescence images directly reveal the ${\rm mm}$-scale spatial distribution of the atomic cloud, they offer no obvious clues about internal properties such as temperature.
We demonstrate that deep learning can extract both the straightforward and the hidden characteristics of these ensembles with high accuracy from real experimental data.

Laser cooling is a foundational technique~\cite{Hansch1975} underpinning virtually all atom-based quantum technologies, including quantum sensors, simulators, and computers.
To harness quantum effects with neutral atoms, these technologies require large collections of ultracold atoms that share the same quantum state of motion and the same internal state (i.e., atomic level).
For some applications, laser cooling alone is sufficient, while in others, it is followed by additional stages of cooling and state purification.
The magneto-optical trap (MOT) is widely used to capture, trap, and cool neutral atom clouds~\cite{Raab1987} ranging in size from tens of billions of atoms down to the single-atom level.
For example, MOTs serve as precursors to today's leading optical lattice clocks~\cite{Ludlow2015}, optical tweezer arrays~\cite{Kaufman2021}, quantum degenerate gases~\cite{Anderson1995}, and much more.

\begin{figure*}[tbh!]
    \centering
    \includegraphics{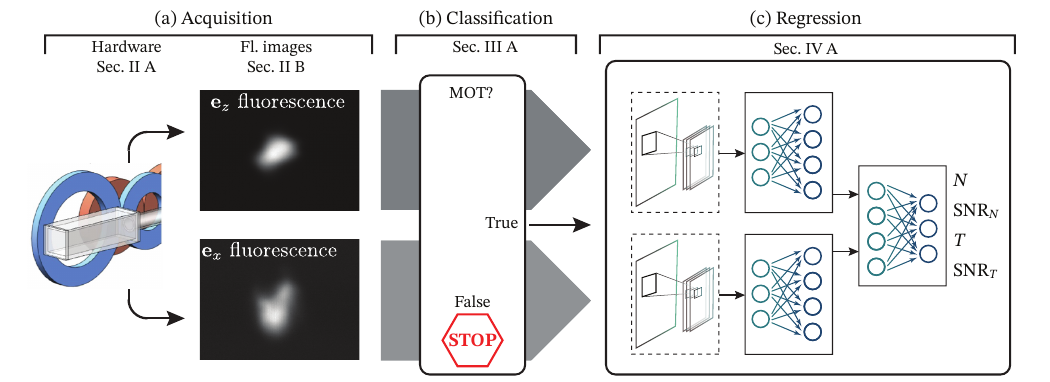}
    \caption{
    Workflow of classification and regression system in operation.
    (a) Acquisition.  $\K39$ atoms are laser cooled and then florescence-imaged along $\ex$ and $\ez$ as described Sec.~\ref{sec:experiment} and \ref{ssec:experimental_sequence}.
    Images for optimal experimental parameters are shown.
    (b) Classification.  After acquisition data is classified.
    (c) Regression.  The data is then passed into a regression model which: first potentially processes the images independently; fuses the data; and predicts atom number $N$ and temperature $T$ along with corresponding signal-to-noise ratios ${\rm SNR}_N$ and ${\rm SNR}_T$.    
    Section~\ref{sec:toolbox} discusses data pre-processing, classification, and regression.
    }
    \label{fig:workflow}
\end{figure*}

Each of these applications, as well as the MOT itself, has been enhanced by ML.
For example, ML has been used to automate the operation of optical atomic clocks in real-world applications~\cite{Roslund2024}.
Furthermore, the control parameters used when loading a MOT~\cite{Tranter2018}, creating optical optical tweezer arrays~\cite{Ren2024}, and producing quantum gases have all been optimized using ML techniques~\cite{Wigley2016,Vendeiro2022}.
Oftentimes, these ML-based optimizers discover unexpected or counterintuitive parameter sequences.
Together these applications demonstrate the breadth of ML's applicability to atom-based quantum science and technology.
Here, we focus on efficiently extracting information from noninvasive images of laser cooled atoms in a MOT.

A MOT operates using an interplay of optical and magnetic forces, and atoms in a MOT constantly scatter light from illuminating lasers in all directions.
Imaging this unavoidable fluorescence, therefore, affords an often-used mechanism for noninvasive monitoring of the trapped cloud.
Although the exact scattering rate of each atom depends on detailed experimental parameters, it is intuitive that the overall amount of scattered light generally increases with atom number; indeed, at low enough density (such that light scattered by one atom is unlikely to be reabsorbed by another), the overall fluorescence is simply proportional to the atom number~\cite{Metcalf1999}.

Except for very simple atoms (internal angular momentum $J=0$ to $J=1$) and very small (negligible rescattering) clouds, neither the total fluorescence nor the cloud's size and shape provide any obvious indication of its temperature, thereby hiding this parameter from standard image-based analyses.
Both number and temperature can be readily measured using invasive techniques such as time-of-flight (TOF)  imaging, in which the atoms are released from the MOT and allowed to ballistically expand for a set time.  
In this way, the spatial distribution after TOF is correlated with the initial velocity distribution, from which the temperature can be estimated~\cite{Lett1988}.
Performing TOF imaging immediately after acquiring fluorescence images thus allows us to generate datasets of fluorescence images obtained noninvasively that are nonetheless labeled by the atom number $N$ and temperature $T$.

In more detail, each element of our dataset (a single ``shot'' of the experiment) includes two fluorescence images captured along two orthogonal axes, together with a reference TOF image.
These individual shots are then collected into ``sets,'' each consisting of $M=5$ shots that differ only in the TOF time, thereby improving the accuracy with which we determine atom number and temperature.
The final dataset consists of $\approx 39\times10^3$ individual shots, whose  $\approx 8\times10^3$ distinct parameters span a nominally balanced portion of parameter space, with contributions ranging from large atomic clouds down to a complete absence of atoms, and everything in-between.
We then train a range of regression models on these datasets to extract $N$ and $T$, as well as corresponding quality metrics, directly from fluorescence images alone.
In operation, new data employing the final trained models follows the overall workflow illustrated in Fig.~\ref{fig:workflow}.

We explored a total of five regression models ranging in sophistication from a trivial model that returned a constant output (serving as our benchmark), to a convolutional neural network (CNN); these models' performance increasing in line with their sophistication.
The final CNN model predicts $N$ with a typical uncertainty of $\pm 4\times10^6$ out of $2\times10^8$ atoms, and $T$ with a typical fractional uncertainty of $\pm0.2$ (see SM for all results).

The paper is structured as follows.
Section~\ref{sec:experiment} describes the experimental setup, relevant parameters, and time-sequence.
Then, Sec.~\ref{sec:data_collection} describes the full dataset including prepossessing and labeling.
Section~\ref{sec:toolbox} introduces the five regression models (CON, LIN, MM, MLP, CNN), data augmentation strategies, and our training procedure.
Results are reported and discussed in Sec.~\ref{sec:results}.
Section~\ref{sec:conclusion} comments on the implications of our findings and provide an outlook.

\section{Experiment}
\label{sec:experiment}

\begin{table*}[tbh!]
    \centering
    \caption{\label{tab:exp_parameters} 
    Dataset labels.
    This table contains the subset of our experimental parameters that are varied between different elements in our dataset (top), the settings that change within each set (middle), and the labels derived by our classification and fitting processes (bottom).}
    \begin{tblr}{colspec = {c c c Q[wd=3.5in]}, hlines}
        Symbol & Label & Approximate range & Description \\ \hline
        
        $V_{\rm cool}$ & {\tt Cooling\_AOM\_Volts} & $0.1$~\si{\volt} to $1.5$~\si{\volt} & Parameter: Voltage of the cooling laser beam's AOM. Controls the intensity of the cooling laser beam. \\ 

         $V_{\rm rep}$ & {\tt Repump\_AOM\_Volts} & $0.4$~\si{\volt} to $1.5$~\si{\volt} & Parameter: Voltage of the repump laser beam's AOM. Controls the intensity of the repump laser beam. \\
        
        $f_{\rm lock}$ & {\tt Cooling\_Lock\_Offset} & $85$~\si{\mega\hertz} to $95$~\si{\mega\hertz} & Parameter: Controls the frequency offset of the cooling laser with respect to the repump laser. \\
        
        $f_{\rm rep}$ & {\tt Repump\_AOM\_Freq} & $74$~\si{\mega\hertz} to $94$~\si{\mega\hertz} & Parameter: Controls the frequency offset of the repump laser beam compared to the repump laser source.  \\
        
        $I_{\rm quad}$ & {\tt MOT\_Quad\_Amps} & $2$~\si{\ampere} to $40$~\si{\ampere} & Parameter: Current of the MOT quadrupole coils. Affects the strength of the magnetic field. \\
        
        $t_{\rm MOT}$ & {\tt MOT\_Loading\_Time} & $100$~\si{\milli\second} to $\num{1800}$~\si{\milli\second} & Parameter: Duration of MOT loading time.  \\
        \hline
        $t_{\rm TOF}$ & {\tt TOF\_Time} & $1~\si{\milli\second}$ to $5~\si{\milli\second}$ & Set variable: Time of flight. \\
        \hline        
        - & {\tt MOT} & [{\tt True}, {\tt False}] & Label: indicates whether there is any identifiable fluorescence signal.\\

        - & {\tt VALID\_SET} & [{\tt True}, {\tt False}] & Label: data for which the labels could not be assigned.\\

        $N$ & {\tt NUM} & $8\times10^6$ to $2\times10^8$ &  Label: Number of atoms. \\

        ${\rm SNR}_N$ & {\tt  NUM\_REL} & $0.1$ to $100$ & Label: Number reliability. \\
        
        $T$ & {\tt TEMP} & $0.5$~\si{\milli\kelvin} to $30$~\si{\milli\kelvin} & Label: Temperature. \\

        ${\rm SNR}_T$ & {\tt TEMP\_REL} & $0.1$ to $100$ & Label: Temperature reliability.
    \end{tblr}
\end{table*}

\begin{figure}
    \centering
    \includegraphics{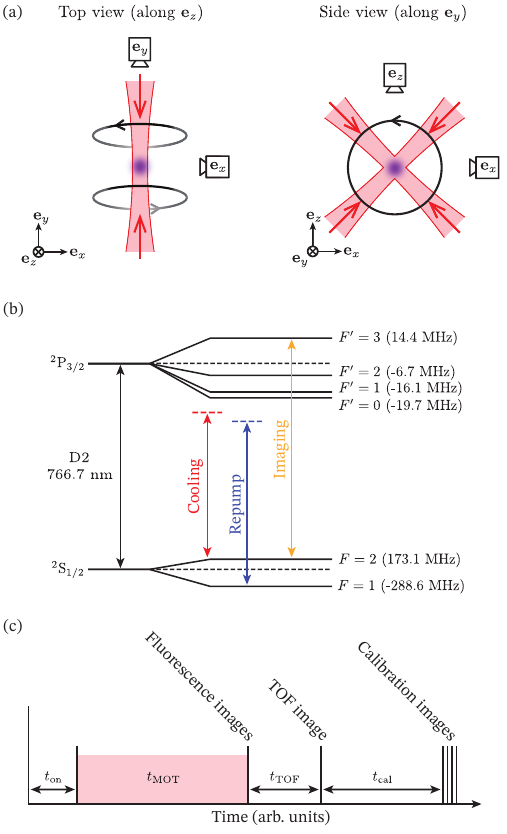}
    \caption{ 
    (a) Schematic of experimental geometry including top (left) and side (right) views. 
    (b) Relevant energy levels for laser cooling and trapping $\K39$ using only the D2 line. 
    (c) Experimental sequence for MOT operation.
    }
    \label{fig:schematic}
\end{figure}

This section briefly describes our experimental hardware, outlines the overall timing and control sequences, and concludes by introducing the overall structure of our data.
The experiment uses laser cooling techniques to create and capture an atomic cloud of $\K39$ atoms using a MOT~\cite{Chu1985,Pritchard1986,Raab1987}.
Our MOT, configured in the standard geometry shown in Fig.~\ref{fig:schematic}(a), relies on radiation pressure from three pairs of counterpropagating laser beams that each include contributions from both ``cooling'' and ``repump'' lasers nearly resonant with the D2 line, along with a quadrupole magnetic field.
While in this work we study a range of parameters, the cooling laser would generally be red-detuned from the $F=2\rightarrow F'=3$ transition, and the repump laser would be tuned near the $F=1\rightarrow F'=2$ transition, [see Fig.~\ref{fig:schematic}(b)].

In this configuration, the lasers' radiation pressure Doppler cools our $\K39$ atoms to a typical temperature of $\approx 2$~\si{\milli\kelvin}, and the inhomogeneous detuning from the quadrupole magnetic field adds confinement. 
The use of cooling and repumping lasers assures that atoms do not accumulate in an unaddressed (dark) ground state, thereby leaving the cooling process.

\subsection{Experimental hardware}
\label{ssec:experimental_hardware}

Our experimental apparatus, shown schematically in Fig.~\ref{fig:schematic}(a), is a standard vapor-fed MOT.
This first captures atoms from the low-velocity tail (below $\approx 30~\si{\meter}/\si{\second}$) of the dilute room temperature $\K39$ vapor in our vacuum system.
These atoms are then cooled and collected, yielding trapped atoms with velocities around $1~\si{\meter}/\si{\second}$.
All of the control parameters that are varied in this study are detailed in Table~\ref{tab:exp_parameters}.

The apparatus makes use of two laser systems: one generates the cooling and imaging laser beams and the other generates the repump beams.
The repump laser system is locked to a potassium reference cell using saturated absorption spectroscopy.
The cooling laser system is then offset locked to the repump using a phase-locked loop circuit giving a tunable frequency offset $f_{\rm lock}$ between these laser systems.
Each final laser beam relies on an acousto-optical modulator (AOM) to provide high bandwidth control of the power (controlled by an external voltage) and to introduce a tunable frequency shift.
In our dataset, the power of both the cooling and repump beams are tuned with control voltages $V_{\rm cool}$ and $V_{\rm rep}$.

The atomic levels relevant to Doppler laser cooling of $\K39$ are shown in Fig.~\ref{fig:schematic}(b); the specific values depicted are representative of our best MOTs.
The vertical black line shows the nominal scale of the D2 transition; the red line shows the cooling laser, red detuned from the $F=2\rightarrow F'=3$ transition; the blue line shows the repump, red detuned from $F=1\rightarrow F'=2$ transition; and lastly the orange line shows the absorption imaging probe frequency resonant with the $F=2\rightarrow F'=3$ transition~\footnote{In the first milliseconds of TOF the magnetic field has not reached zero, so we optimized the frequency offset at each $t_{\rm TOF}$.}.
(Unlike atoms such as $\Rb87$ or $\Na23$, in $\K39$, the excited state hyperfine splitting is poorly resolved, being close to the $\approx 6\ {\rm MHz}$ transition linewidth.  In this configuration, the repump laser contributes significantly to the cooling and trapping forces.)

In assembling our data set, the repump AOM provides a variable frequency $f_{\rm rep}$, but the cooling AOM frequency is fixed.
Instead, the offset lock provides frequency tuning to the cooling laser system, and does so over a wider frequency range than is possible with an AOM.

To increase the geometric stability of our experiment, all of the laser light is injected into optical fiber after being conditioned by the AOMs.
The probe laser travels in a conventional single mode fiber, while the cooling and repump light is combined by an evanescent wave fiber splitter-combiner that distributes the optical power equally into the requisite six beams.

In addition to these optical fields, MOT operation requires a quadrupole magnetic field.
We generate this field with a pair of copper coils arranged in an anti-Helmholtz configuration, each carrying the same current $I_{\rm quad}$.
Additional coils (not shown) compensate for the ambient background magnetic field.

\begin{table*}[tbh!]
    \centering
    \caption{\label{tab:imaging} 
    Imaging parameters.
    This table describes the relevant information of our three imaging systems~\cite{Equipment}.
    The $\ex$ and $\ez$ magnifications differ due to geometric constraints of the optical layout and the available lens selection.
    }
    \begin{tabular}{ c c c c c c c }
        \hline\hline
        Camera name & Model & Resolution & Magnification & Magnified pixel size & Measurement type \\ \hline
        
        $\ex$-camera & Mako G-030B & $644\times484$ & 0.5 & $14.8\ \mu {\rm m}$ & fluorescence imaging \\ 
        
        $\ey$-camera & Mako G-131B & $1280\times1024$ & 0.333 & $15.9\ \mu {\rm m}$ & absorption imaging \\
        
        $\ez$-camera & Mako G-030B &  $644\times484$ & 0.375 & $19.7\ \mu {\rm m}$ & fluorescence imaging \\ 
        
        \hline\hline
    \end{tabular}
\end{table*}

The atomic ensemble is imaged along the three Cartesian axes ($\ex$, $\ey$, and $\ez$) using independent two-lens Keplerian microscopes.
The images are captured on complementary metal oxide semiconductor (CMOS) cameras, each labeled by its imaging axes, for example the ``$\ex$-camera.''
The important properties of these imaging systems are detailed in Table~\ref{tab:imaging}.
Because we are imaging large ${\rm mm}$-scale objects, our images are demagnified with effective pixel sizes that are larger than the $~\approx 6\ \mu{\rm m}$ diffraction limit of these imaging systems.

\subsection{Experimental sequence}
\label{ssec:experimental_sequence}

\begin{figure}[!tb]
    \centering
    \includegraphics{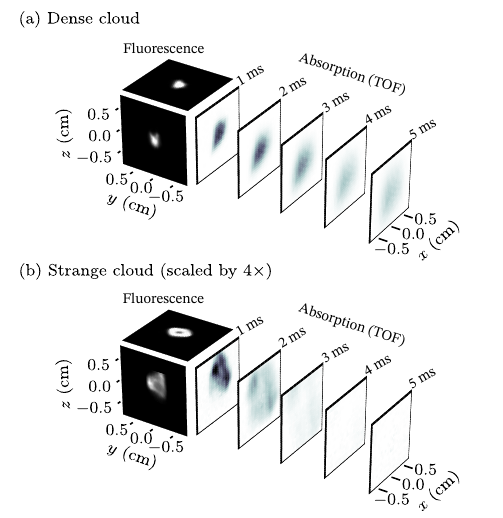}
    \caption{Comparison between compact and diffuse clouds.
    (a) and (b) correspond to ({\tt BATCH 00}, {\tt SET 150}) and ({\tt BATCH 03}, {\tt SET 995}) respectively.
    Note that the signal in (b) is scaled by $4\times$ to make this lower quality MOT visible.
    }
    \label{fig:TypicalData}
\end{figure}

This section outlines the time-sequence of a single experimental shot yielding an elementary unit of a dataset.
This sequence (potentially) generates a cloud of laser cooled atoms from a specific set of experimental parameters.
Each such shot follows a predefined sequence of events organized into stages of: initialization, MOT loading, fluorescence imaging, TOF imaging, and calibration [see Fig.~\ref{fig:schematic}(c)].
Figure~\ref{fig:TypicalData} shows fluorescence and TOF data, for dense and compact (top) as well as more representative clouds (bottom).

{\it Initialization}---Prior to MOT loading, we allow for a period of hardware equilibration of duration $t_{\rm on} = 5$~\si{\milli\second}.
During this time the parameters {\tt Cooling\_Lock\_Offset}, {\tt Repump\_AOM\_Freq}, {\tt MOT\_Quad\_Amps}, {\tt Cooling\_AOM\_Volts} and {\tt Repump\_AOM\_Volts}) are set, with the lasers mechanically blocked by shutters just prior to the entering optical fibers.

{\it MOT loading}---Cooling and trapping is then initialized by abruptly opening the shutters.
This stage has a duration $t_{\rm MOT}$, equal to the {\tt MOT\_Loading\_Time} parameter in Table \ref{tab:exp_parameters}.

{\it Fluorescence imaging}---Immediately following MOT loading, the $\ex$- and $\ez$-cameras acquire their respective fluorescence images.

{\it TOF imaging}---The magnetic and optical fields are then removed, thereby freeing the atoms from the trap, after which time they undergo TOF evolution for a duration $t_{\rm TOF}$.
The resulting two-dimensional (2D) column density $\rho_{ij}$ in each pixel (labeled by $i,j$) is measured via absorption imaging, a process that in essence detects the shadow cast by the atom cloud in a probe laser.
This image is acquired by pulsing on the probe laser (traveling along $\ey$) for $10\ \mu{\rm s}$, and an auxiliary repump beam (traveling along $\ez$) starting $20\ \mu{\rm s}$ prior to the probe pulse.
The $\ey$-camera then measures the shadowed probe.

{\it Calibration}---The raw fluorescence and absorption images require additional reference data to mitigate the effect of background light as well as calibrate the un-shadowed probe profile (see Ref.~\onlinecite{SM} for a description of these reference frames).
After TOF imaging, these additional images are acquired, adding a time $t_{\rm cal} = 411\ {\rm ms} + 2 t_{\rm TOF}$ to each shot.

\section{Data}
\label{sec:data_collection}

We collected data under a wide range of experimental conditions to produce a diverse dataset for model training.
This was achieved by varying the six experimental parameters in Table~\ref{tab:exp_parameters}.
The values of these parameters were sampled so as to generate an approximately balanced dataset.

The dataset consists of $14$ batches, each ranging in size from $14$ to \num{1000} sets of shots.
Each set contains $M=5$ shots with identical MOT parameters, except that $t_{\rm TOF}$ samples the set $\{1,2,3,4,5\}\ \si{\milli\second}$.
Each shot yields two fluorescence images taken at the end of MOT loading and one absorption image taken after a subsequent $t_{\rm TOF}$ expansion.
In total, the dataset contains \num{38915} shots, as described in detail in Ref.~\onlinecite{DAmato2025}.
Each data file contains the experimental parameters used to generate the shot, as well as all image frames described in Sec.~\ref{ssec:experimental_hardware} and \ref{ssec:experimental_sequence}.

\subsection{Labeling strategy}
\label{sec:labeling}

Each shot in the dataset is assigned the six labels shown in Table~\ref{tab:exp_parameters}; these labels are all assigned at the set level.
Here we describe our strategy for generating these labels in the order they are assigned in our labeling pipeline.

\vspace{3pt}

{\tt MOT}: This label identifies data with overall fluorescence in excess of the background noise level and is either {\tt True} or {\tt False}.
Because our data is organized into sets sharing the same parameters, the {\tt MOT} label for a specific shot is assigned {\tt True} only if every image in its set has a detectable fluorescence signal.

\begin{figure}[tb!]
    \includegraphics{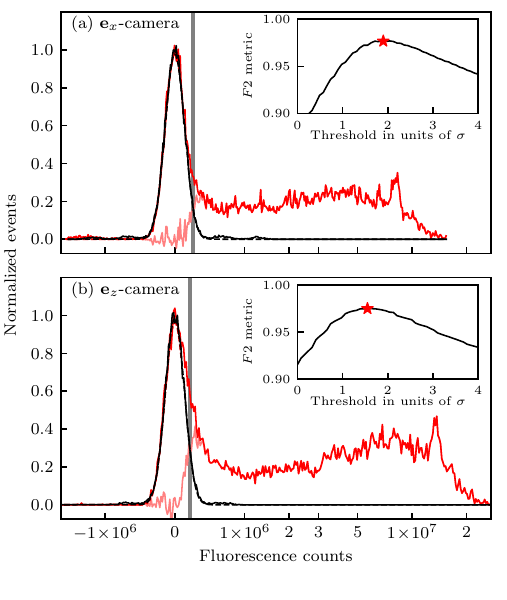}
    \caption{Normalized fluorescence histograms.
    Panels (a) and (b) derive from  the $\ex$- and $\ez$-cameras respectively.
    Horizontal axes are arcsinh-scaled, a symmetric scaling which interpolates between linear for small arguments and logarithmic for large arguments.
    Solid black curves, obtained with no atoms present, are plotted along with Gaussian fits (dashed black curves).
    Red curves describe the complete dataset, while the pink curves are the difference between the complete and the no-atoms histograms.
    The vertical gray lines represents the threshold below which data is assigned {\tt MOT} = {\tt False}.
    Insets: $F2$ metric as a function of threshold with operating point marked in red.
    }
    \label{fig:threshold}
    
\end{figure}


We separately determine the background noise level for the $\ex$- and $\ez$-cameras from fluorescence frames taken with no atoms present.
The pixel values in each such mock fluorescence image are summed to obtain an overall signal shown by the black curves in Fig.~\ref{fig:threshold}; Gaussian fits yield a root mean square (RMS) width of ${\rm RMS} = 1.22\times10^5$ counts for both cameras.
These curves are then plotted along with corresponding histograms of the summed fluorescence signal $S_x$ and $S_z$ from true fluorescence images (red), showing good agreement for the nonphysical negative portion of the fluorescence signal.
We estimate the true distribution by subtracting the background distributions (pink curves).

Together these distributions allow us to compute the F2 metric for any proposed threshold (inset). 
In practice, we select thresholds of $1.9\times{\rm RMS}$ and $1.5\times{\rm RMS}$ (horizontal gray lines) for the $\ex$- and $\ez$-cameras respectively, where the F2 metrics achieve their maximum values of $0.977$ and $0.975$.

\vspace{3pt}

{\tt NUM} and {\tt TEMP}: These labels, describing the cloud's atom number and temperature, are real-valued, and by default are assigned {\tt NaN} when ${\tt MOT} = {\tt False}$.
Both quantities can be obtained from the spatial distribution of atoms expanding during TOF.
The integrated distribution directly yields atom number, while the evolution of the distribution over time provides access to the velocity distribution and, therefore, an effective mean thermal energy~\cite{Lett1988}.

In order to make this labeling stage fast and reliable, we make the simplifying assumptions that both the velocity distribution and the initial density distribution are Gaussian (this process is further accelerated by down-sampling the 2D column density images from $\num{1024}\times\num{1280}$ to $51 \times 64$ pixels).
The resulting model distribution is Gaussian at every $t_{\rm TOF}$, with density
\begin{align}
\rho(x,z) &=  \frac{N}{2\pi w_x w_z} \exp\left[-\frac{1}{2}\left(\frac{x^2}{w_x^2} + \frac{z^2}{w_z^2}\right)\right];
\end{align}
because our TOF data is imaged along the $\ey$ axis, we cannot access the distribution along $\ey$.
This expression is in terms of the TOF-expanded RMS widths
\begin{equation}
    \label{eq:variance_thermal_expansion}
    w_{x,z}^2(t_{\rm TOF}) = w_{x,z}^2(0) + \frac{\kB T}{m}t_{{\rm TOF}}^2,
\end{equation}
the initial widths $w_{x,z}(0)$, the atom number $N$, the temperature $T$, the atomic mass $m$, and Boltzmann's constant $\kB$.

For each set, we perform a joint fit to its $M$ shots (each at a different $t_{\rm TOF}$), yielding a single number and temperature.
In practice, we find that the distribution expands at different rates along $\ex$ and $\ez$, giving separate values $T_x$ and $T_z$, which we average to yield the final temperature label.
Ref.~\onlinecite{SM} details the fitting process.

\vspace{3pt}

{\tt NUM\_REL} and {\tt TEMP\_REL}: These labels describe the reliability of the number and temperature labels, are real-valued, and by default are assigned {\tt NaN} when ${\tt MOT} = {\tt False}$.
As seen in Fig.~\ref{fig:TypicalData}, the observed TOF density distributions can be far from Gaussian; therefore, the fit uncertainties, $\Delta N$ and $\Delta T$, reported by our Gaussian model cannot be simply interpreted as statistical uncertainties.
Instead, they reflect an uncalibrated combination of statistical uncertainties and systematic artifacts that we use to define heuristic reliability indices ${\tt NUM\_REL} = | N / \Delta N | \equiv {\rm SNR}_N$ and ${\tt TEMP\_REL} = |T / \Delta T| \equiv {\rm SNR}_T$ that can loosely be interpreted as signal-to-noise (SNR) ratios.

\vspace{3pt}

{\tt  VALID\_SET}: This boolean label is assigned {\tt True} unless the set is rendered invalid for technical reasons such as: one or more images were not acquired, or a labeling fit failed to converge.

\subsection{Data organization}
\label{sec:data_organization}

After labeling, we set aside all of batch 6 (684 sets, about $10\ \%$ of the complete dataset) as an out-of-distribution test set, and a randomly selected $10\ \%$ of the remaining 13 batches for a traditional in-distribution test set.
These test sets were completely excluded from the training process: their contents were never used for training, nor were they used to guide the model development process.
All training and validation therefore only used the $\approx80\ \%$ of the data that were not part of either test set; further subdivision of the data, for example for $N$-fold cross-validation, is discussed in the context of training in Sec.~\ref{ssec:training}.

\section{Machine learning toolbox}
\label{sec:toolbox}

Here we describe the mechanism by which fluorescence image data from the $\ex$- and $\ez$-cameras are processed as they move through our ML analysis pipeline.
Prior to any further processing: (1) the {\tt MOT} label is determined as described in Sec.~\ref{sec:labeling} by separately comparing the summed fluorescence counts from the $\ex$- and $\ez$-images to the predetermined thresholds; (2) the pixel values are divided by \num{4096}, normalizing them to the maximum signal of our 12-bit CCD cameras; and (3) the resolution of both florescence images are reduced from $644\times484$ to $64\times48$.
Processing then terminates for data with ${\tt MOT} = {\tt False}$.

\subsection{Regression}\label{ssec:ML_models}

This section describes the regression models employed to predict the number, temperature, and the associated reliabilities using only the fluorescence images.
The least sophisticated of these ``models'' is a constant (CON) function that returns the same output irrespective of its input; this defines the baseline performance level to which all other models are compared.
We then progress to a simple linear function (LIN) of the summed fluorescence counts, to a single-layer linear model (i.e., matrix multiplication, MM), then to a multi-layer perceptron (MLP), and culminate with a convolutional neural network (CNN).
These models approximately recover $N$, ${\rm SNR}_N$, $T$, and ${\rm SNR}_T$, with performance increasing in line with their sophistication (the parameter count of all models is tabulated in Ref.~\onlinecite{SM}).

All of these models are optimized with respect to the overall loss function $L^2=D^{-1}\sum\ell^2$, averaged over a $K$-element data set.
Each member of the dataset has an individual loss
\begin{align}
\ell^2 =& \ \SNR_N^2 \left\{\left[1-\frac{N'}{N}\right]^2 + \left[1-\frac{\SNR_N'}{\SNR_N}\right]^2\right\} + \nonumber \\
& \ \SNR_T^2 \left\{\left[1-\frac{T'}{T}\right]^2 + \left[1-\frac{\SNR_T'}{\SNR_T}\right]^2\right\}, \label{eq:loss}
\end{align}
derived from model predictions of the physical parameters $N'$ and $T'$ and their reliability indices $\SNR_N'$ and $\SNR_T'$.
This loss function minimizes the fractional uncertainties in each quantity, weighted by the SNR predicted by the fit, i.e., our reliability label.

This is equivalent to the standard weighted least squares loss function, where the error terms such as $(N-N')^2$ are weighted by $\Delta N^2$.
Therefore, $\ell^2$ can be interpreted as the L2 norm of the loss vector
\begin{align}
\bm{\ell} \! &=\! \left[\frac{N\!-\!N'}{\Delta N}, {\rm SNR}_N\!-\!{\rm SNR}_N',\frac{T\!-\!T'}{\Delta T}, {\rm SNR}_T\!-\!{\rm SNR}_T'\right].\label{eq:residual_vector}
\end{align}
Because $\bm{\ell}$ consists of ratios of like quantities, every component is of nominally comparable scale; we therefore did not require additional relative weighting hyper-parameters.

\vspace{3pt}\noindent
\textbf{Constant output}.
The CON model is the absolute minimal case that returns the same values irrespective of its input and therefore has 4 ``trainable'' parameters, one per regression variable.
The simple form of the loss function allows us to express these parameters in closed form.
For a dataset with $K$ elements, and arbitrary labels $\big\{A_k\big\}_{k=1}^K$ and $\big\{\SNR_{A,k}\big\}_{k=1}^K$, the corresponding contribution to Eq.~\eqref{eq:loss} is minimized by
\begin{align}
A' &= \frac{\sum_{k=1}^K A_k \Delta A_k^{-2}}{\sum_{k=1}^K \Delta A_k^{-2} },
\end{align}
where $\Delta A_k^{-2} = (\SNR_{A,k} / A_k)^2$ serve as statistical weight factors.

\vspace{3pt}\noindent
\textbf{Linear regression}.
We now incrementally increase complexity with the LIN model, a linear function of the summed fluorescence counts $S_x$ and $S_z$.
For example, atom number is predicted by
\begin{align}
N' &= a_x S_x + a_z S_z + b,
\end{align}
with learnable slopes $a_x$ and $a_z$ and offset $b$.
Thus the overall model for our 4 regression variables has 12 parameters.

\vspace{3pt}\noindent
\textbf{Matrix multiplication}.
We continue to increase complexity by turning to MM, the most general linear model: an offset matrix product (equivalent to a single fully connected layer with bias and linear activation).
To do so, we address the multi-input nature of the dataset with early fusion~\cite{Baltrusaitis2019} in which we flatten both images into one-dimensional (1D) vectors and concatenate them.
In terms of the resulting data vector $d_j$, with dimension $D = 2\times(64\times48)=\num{6144}$, this linear model predicts a four-dimensional (4D) vector
\begin{align}
p_i' &= \sum_j A_{ij} d_j + b_i,\label{eq:MM}
\end{align}
where the linear transform is encoded by the $4\times D$ matrix $A_{ij}$ and 4D offset vector $b_i$, yielding a total of $4D + 4$ parameters.

\vspace{3pt}\noindent
\textbf{Multi-Layer perceptron}.
The MLP is a bona-fide deep learning model; our implementation employs an intermediate fusion approach~\cite{Baltrusaitis2019} to reduce the parameter count.
As schematically illustrated in Fig.~\ref{fig:workflow}, the images are individually flattened and propagated through a sequence of independent fully connected layers whose output vectors are then concatenated (fused).
The fused vector is then passed through a series of fully connected layers.
The MLP is distinguished from MM in that each layer consists of matrix multiplication followed by a non-linear activation function (in our case leaky ReLU~\cite{Maas2013}).
Without an activation function, an arbitrary number of linear layers can always be reduced into a single matrix product as in Eq.~\eqref{eq:MM}.

\vspace{3pt}\noindent
\textbf{Convolutional neural network}.
CNN architectures are effective in identifying spatial patterns, making them well-suited for image analysis; each convolutional layer convolves its input with one or more learned feature kernels, applies a non-linear activation function, and pools the outputs.
Convolutional layers identify features in a translationally invariant way, and in conventional implementations such as ours, successively down-sample the image resolution.

Similar to the MLP model, our CNN also employs the intermediate fusion approach, here with two parallel series of convolutional layers.
The outputs of these layers are flattened, concatenated into a single vector, and passed through a series of fully connected layers yielding a 4D output.
These fully connected stages are not translationally invariant, implying that our overall ``CNN'' model can also learn information related to where features reside in the images.

\subsection{Data augmentation}
\label{ssec:data_augmentation}

We employ data augmentation via geometric transformations to increase the robustness of our models to out-of-distribution data and the stability of the training.
Both fluorescence images could be: unaltered (U), randomly reflected (R), randomly translated (T), or both (RT).
These transformations are physically realistic, as if the three-dimensional (3D) atom cloud giving rise to the images has itself been reflected and/or translated in 3D space.
Augmentations are re-randomized for each training epoch.

\begin{figure*}[tb!]
    \centering
    \includegraphics{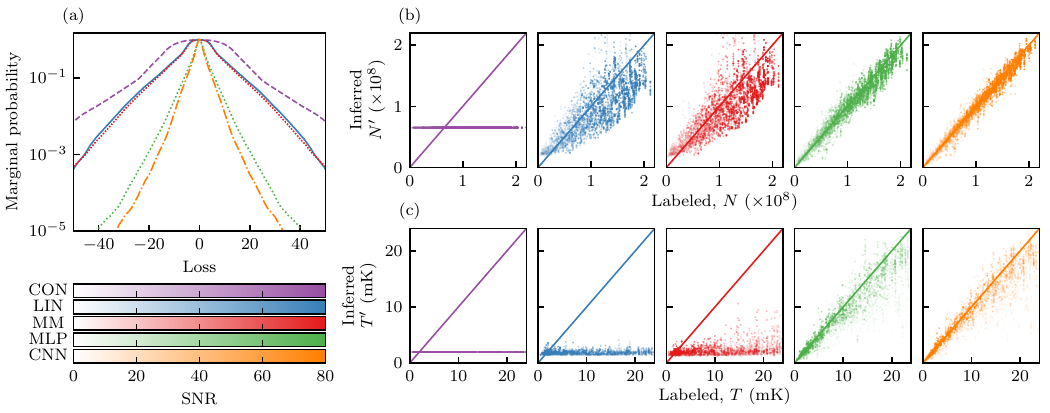}
    \caption{
    Inference for the CON (dashed purple), LIN (solid blue), MM (fine-dotted red), MLP (dot-dotted green) and CNN (dash-dotted orange) models.
    These models are trained using RT augmentation, and compared to the in-distribution test set with RT augmentation (configured to give a ten-fold increase in test data).
    To focus the effect of the test distribution's statistics, we evaluate a randomly selected models from those trained in the 10-fold cross validation.
    (a) marginal distributions of residual vectors $\bm{\ell}$.
    (b) and (c) number and temperature inference respectively.
    Points are shaded according to their SNR (color bars) and lines of slope one indicate the ideal behavior.
    }
    \label{fig:summary_results}
\end{figure*}

\vspace{3pt}\noindent
\textbf{Reflection}.
Our physical system is symmetric under reflection about planes normal to the three Cartesian axes, generated by operators $R_{x,y,z}$.
As a result, our reflection-augmentation is implemented by elements of the group generated by these (commuting) operators $\left\{I, R_x, R_y, R_z, R_{xy}, R_{yz}, R_{zx}, R_{xyz} \right\}$; where $I$ is the identity; each group element $R$ obeys $R^2 = I$; and for example, $R_{xy} \equiv R_xR_y$.
We do not include other symmetry allowed operations such as rotations about $\ez$ because there is insufficient information in our two fluorescence images to generate such data (such operations correspond to observations of the cloud along arbitrary axes in the $\ex$-$\ey$ plane). 
When this augmentation is employed, a different randomly selected operator (including the identity) is applied to each element of the training dataset.

\vspace{3pt}\noindent
\textbf{Translation}.
Although our system is not translationally invariant, the observed position of the laser cooled atoms on the sensors results from the manual alignment of the imaging systems, and is prone to small changes when the system is reconfigured.
In addition, variability in the alignment of the MOT lasers as well as stray magnetic fields lead to translations of the cloud.
We therefore augment via 3D translations of the atomic cloud, constrained so that the integrated signal is reduced by no more than $10~
\%$; this confines translations to a cube of side $6~{\rm mm}$.

When this augmentation is active, we randomly select displacement vectors where each component is uniformly distributed within the allowed domain. 

\vspace{3pt}\noindent
\textbf{Reflection and Translation}.
Lastly, we combine both forms of augmentation by implementing a reflection followed by a translation.
Both the reflection plane and the translation vector are drawn at random from the distributions described above.

\subsection{Training}
\label{ssec:training}

All models are implemented in PyTorch~\cite{Paszke2019} and are fully detailed in Ref.~\onlinecite{SM}.
Of these models, only the MLP and CNN have tunable hyperparameters (e.g., number of layers, layer size, kernel size, etc.), and we selected their values heuristically to obtain performant outcomes.
It is likely that careful hyperparameter optimization will improve the performance of fully trained models.

Every model is then separately trained on data augmented using each strategy discussed in Sec.~\ref{ssec:data_augmentation}.

We employ mini-batch gradient descent using the Adam optimizer and an adaptive learning rate reduction scheme with a base learning rate of $10^{-4}$.
For consistency, all models are trained for \num{4000} epochs; however, we observe that the number of epochs required before saturation varies drastically depending on the model type.
For example, the LIN model training saturates after about 20 epochs, while the MLP improvement slows around \num{1000} epochs.
In all cases, the validation data confirms that overfitting is not occurring.

\section{Results and discussion}
\label{sec:results}


\begin{table*}[!th]
    \centering
    \caption{Summary of results.
    Models trained with either U or RT augmentation, and evaluated on RT-augmented in-distribution test data.
    Augmenting the test data enlarged the dataset by $\times 10$.
    The quoted values reflect the average and standard deviation of the model outcomes based on a 10-fold cross-validation.
    }
    \label{tab:summary_results}
    \vspace{0.6mm}
    \begin{tabular}{ccccccccccc}
\hline\hline 
 & \multicolumn{2}{c}{CON} & \multicolumn{2}{c}{LIN} & \multicolumn{2}{c}{MM} & \multicolumn{2}{c}{MLP} & \multicolumn{2}{c}{CNN} \\
&\multicolumn{2}{c}{--} & U & RT & U & RT & U & RT & U & RT \\
\midrule
$\FVU$ & \multicolumn{2}{c}{1 (exact)} & 0.70(2) & 0.563(1) & 3.3(2)$\times 10^{2}$ & 0.522(3) & -1.2(9)$\times 10^{2}$ & 0.078(4) & 3.8(6) & 0.052(3) \\
$\Delta N \times 10^{-6}$ & \multicolumn{2}{c}{43.791(3)} & 27.7(2) & 23.82(5) & 285(9) & 23.76(7) & 5(1)$\times 10^{1}$ & 10.4(4) & 48(2) & 7.5(2) \\
$\mathcal{F}_T$ & \multicolumn{2}{c}{0.75542(6)} & 0.758(3) & 0.7484(4) & 11.9(2) & 0.7302(9) & 7(3) & 0.39(3) & 1.4(1) & 0.31(3) \\
\hline\hline
\end{tabular}
\end{table*}


This section cross-compares performance for each model architecture and augmentation.
Training is performed via $10$-fold cross-validation (with randomly selected folds), yielding 10 trained models for every architecture/augmentation combination.
Unless otherwise stated, we report the average result of all 10 models with uncertainties given by the standard deviation.

\subsection{Overview of results}
Figure~\ref{fig:summary_results} provides a high-level summary of results for models trained and tested with RT-augmentation.
Panel (a) shows the marginal distributions of residual vectors $\bm{\ell}$ [Eq.~\eqref{eq:residual_vector}], obtained by projecting the underlying 4D distribution onto all possible 1D axes.
This construction yields unbiased, axis-independent 1D distributions whose widths systematically decrease as model sophistication increases from CON to CNN.
We quantify the relative change in width using the fraction of variance unexplained ($\FVU = 1-R^2$ in terms of the weighted coefficient of determination $R^2$), which is directly related to our loss function by $FVU = L^2/L^2_{\rm C}$.
Here $L^2$ is the total loss for a specific model, and $L^2_{\rm C}$ is that of the trivial CON model (thus $\FVU\equiv1$ for the CON model).
Thus, $\FVU$ measures the fraction of the variance that is unexplained by the model relative to the constant baseline, a quantification of the information learned by the model.

Next, Figs.~\ref{fig:summary_results}(b) and (c) plot the inferred number and temperature as a function of the corresponding labels, with solid lines marking the desired one-to-one correspondence.
Markers are colored according to model architecture, with intensity given by the SNR.
For the inferred number in (b), the deviation from the lines reduces for successive model architectures (from left to right).
It is not surprising that even the LIN and MM models show a modest degree of correlation because, everything else being equal, the overall amount of light scattered during laser cooling increases with atom number.
This contrasts with temperature inference in panel (c), which shows that the LIN and MM models hardly improve upon the CON model.
Only the MLP and CNN models have significant predictive power.

For these MLP and CNN architectures, the error in number is largely independent of $N$, while that of temperature is proportional to $T$.
We therefore report performance in terms of: $\FVU$, the RMS number error $\Delta N$, and the RMS fractional temperature error ${\mathcal F}_T$.
The key numerical results are summarized in Table~\ref{tab:summary_results}, highlighting models trained using U- and RT-augmentation and tested against RT-augmented data.
For the brave of heart, complete results are tabulated in Ref.~\onlinecite{SM}.


\begin{figure*}[t]
   \centering
   \includegraphics{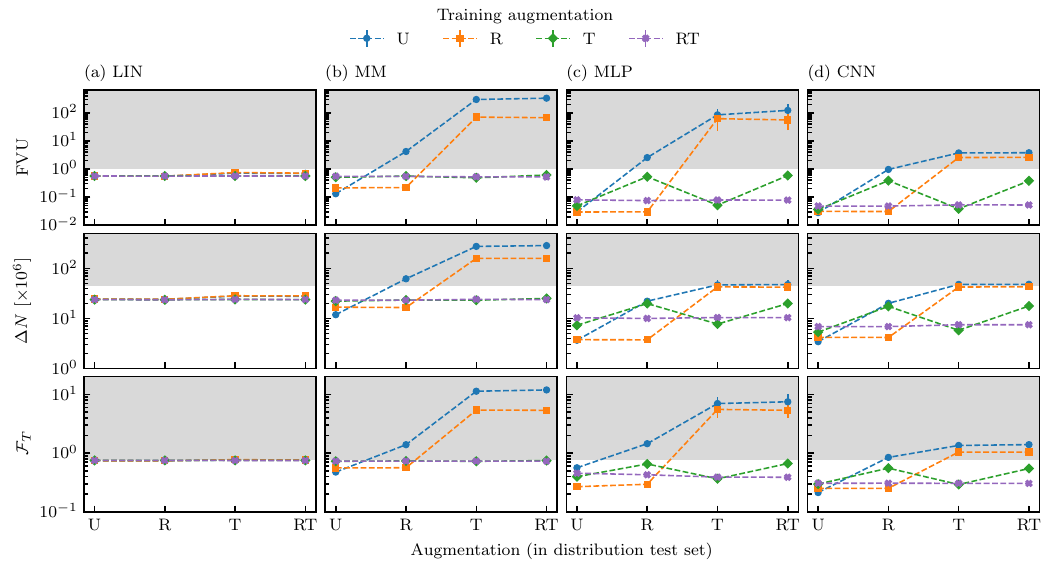}
   \caption{
   Model performance on in-distribution test data.
   Columns (a) - (d) denote results for LIN, MM, MLP, and CNN, respectively; rows plot the three performance metrics $\FVU$, $\Delta N$, and ${\mathcal F}_T$; and the horizontal axes mark the different testing augmentations.
   The gray shaded regions indicate performance below that of the CON model.
   Lastly, the choice of training augmentation is denoted by the marker color: U (blue), R (orange), T (green), and RT (purple).
   Each point reflects the average from a 10-fold cross-validation, with error bars indicating the standard deviation.} 
\label{fig:results}
\end{figure*}


\subsection{Impact of model architecture and augmentation}

We continue with a more detailed discussion of performance as it relates to the choice of both model architecture and augmentation.
Figure~\ref{fig:results} plots $\FVU$, $\Delta N$, and ${\mathcal F}_T$, for which smaller values indicate improved performance.
Model architectures are presented in columns, testing augmentations are indicated on horizontal axes, and training augmentations are distinguished by color.
Because the three performance metrics have similar overall trends, we focus our discussion on $\FVU$ and reserve discussion of $\Delta N$ and ${\mathcal F}_T$ for cases when they exhibit noteworthy behavior.

\vspace{3pt}\noindent
\textbf{Constant output}.
The CON model sets the baseline performance; to facilitate comparison to other models, its outcomes are delineated by the bottom of the gray shaded area in every panel.
The outcomes, $\FVU = 1$ (by definition), $\Delta N \approx 44\times10^6$, and ${\mathcal F}_T \approx 0.76$ are properties of the distribution of labels and therefore independent of augmentation.

\vspace{3pt}\noindent
\textbf{Linear regression}.
Figure~\ref{fig:results}(a) shows that the LIN model's performance is nearly independent of augmentation, degrading only slightly for T and RT. 
This is because the translation operation (and to a much lesser degree, reflection) can shift a small fraction of the fluorescence signal out of the image.
The overall $\FVU$ values are slightly below one, an improvement compared to CON, though most of this improvement results from the reduction in $\Delta N$.
The LIN model has essentially the same accuracy as the CON model for ${\mathcal F}_T$, showing that total fluorescence has no obvious information about temperature.
These observations signify the slight correlation in Fig.~\ref{fig:summary_results}(b) or number, and the lack thereof in (c) for temperature. 

\vspace{3pt}\noindent
\textbf{Matrix multiplication}.
As seen in Fig.~\ref{fig:results}(b) for the MM model, certain combinations of training and testing augmentation yield performance exceeding that of the LIN model.
Specifically, U-training with U-testing and R-training with R- or U-testing are markedly improved.
However, U- and R-training dramatically worsens performance in all other test cases, underperforming even the simple CON model. 
By contrast, T- and RT-trained models have indistinguishable performance that is robust across all test cases but has reduced to that of the LIN model in Fig.~\ref{fig:results}(a).

Together, these results prove that the spatial structure present in fluorescence images (as in Fig.~\ref{fig:TypicalData}) contains information relating to both atom number and temperature; we comment further on this in Sec.~\ref{sec:conclusion}.
Because the only translationally invariant MM-kernels consist of constant entries, this added information is erased by T- and RT-augmentation.
This furthermore suggests that spatial patterns present in MM-kernels obtained for U- and R-training are incompatible with translated data, leading to the worsened performance discussed above. 

\vspace{3pt}\noindent
\textbf{Multi-layer perceptron}.
The MLP results in Fig.~\ref{fig:results}(c) display further improvement, but with the same overall dependency on augmentation as the MM model.
A key difference is that models trained with T-augmentation no longer perform well on R- and RT-augmented test data, indicating that these datasets contain learnable information violating the expected reflection symmetry.

Mirroring the MM results, RT-trained MLP models are robust, with performance that is essentially independent of test-augmentation.
MLP models dramatically improve performance with $\FVU$, $\Delta N$, and ${\mathcal F}_T$ all exceeding the best-case MM models.
Thus, unlike the linear MM model, the non-linear activation functions between the MLP layers enable information regarding spatial structure to be retained even with T- and RT-augmented training.

A final noteworthy observation is that for MLP models trained and tested without augmentation, ${\mathcal F}_T$ is the worst (largest) outcome of any training augmentation [Fig.~\ref{fig:results}(c), bottom].
Nevertheless for U-training, the overall loss ($\propto \FVU$) for U-testing does not exceed that of the other test configurations.
While may seem surprising, it results from our optimization of uncertainty-weighted, not absolute, residuals.

\vspace{3pt}\noindent
\textbf{Convolutional neural network.}
Figure~\ref{fig:results}(d) concludes with our CNN models; as compared to the MLP models, these have a similar augmentation dependence but improved performance.
The CNN architecture consists of a set of translationally invariant convolutional input layers followed by densely connected output layers.
Together, these features yield a smaller decrease in performance for U- and R-trained models evaluated on T- and RT-test data, while still retaining some information regarding absolute position, making U- and T-augmentation inequivalent.
The RT-trained CNN achieves a performance of $\FVU \approx 0.05$ for all augmentations: the best of our models.

\subsection{Out-of-distribution data}
\label{sec:results_out_distribution}

\begin{table*}[tb!]
\centering
    \caption{Out of distribution data.
    Models were trained with either U- or RT-augmentation, and we report the overall loss computed for U-augmented in- and out-of-distribution test datasets.
    Values reflect the average of the model outcomes based on a 10-fold cross-validation, and uncertainties represent the sample standard deviation of the $\approx3\times10^3$ element test datasets evaluated across the 10-folds.  
    }
    \label{tab:out_distribution_results2}
\begin{tabular}{cccccccccc}
\hline \hline
& CON & \multicolumn{2}{c}{LIN} & \multicolumn{2}{c}{MM} & \multicolumn{2}{c}{MLP} & \multicolumn{2}{c}{CNN} \\
& - & U & RT & U & RT & U & RT & U & RT \\
\midrule
$L_{\rm in}$ & 691(6) & 387(4) & 389(4) & 91(1) & 381(4) & 20.5(3) & 55(1) & 19.4(3) & 33.3(4) \\
$L_{\rm out}$ & 721(6) & 380(3) & 388(3) & 106(1) & 374(4) & 57.6(6) & 103(2) & 63(1) & 72(1) \\
\hline
$(L_{\rm out} - L_{\rm in}) / L_{\rm in}$ & 0.04(1) & -0.02(1) & 0.00(1) & 0.16(2) & -0.02(2) & 1.81(4) & 0.87(3) & 2.24(5) & 1.17(3) \\
\hline\hline
\end{tabular}
\end{table*}

Our augmentation process was designed to simulate variability that could, in principle, result from drifting external experimental parameters such as ambient magnetic fields, laser powers, or optical alignment.
The out-of-distribution test data set discussed in Sec.~\ref{sec:data_organization} allows us to access the impact of augmentation.
Because the out-of-distribution data was unseen until final training was complete, all decisions regarding our physically motivated augmentations were uninformed by out-of-distribution performance.

The data in Table~\ref{tab:out_distribution_results2} compares the overall loss $L$ of models trained with either U-augmentation or RT-augmentation evaluated on both in- and out-of-distribution U-augmented data.
Note that degraded performance results in an increase in $L$.
First, performance decreases by a modest but statistically meaningful amount for the CON model, implying that the out-of-distribution data is drawn from a distinguishably distinct distribution.
For the LIN model, this decrease vanishes for either training augmentation, both of which show performance differences consistent with zero.

The remaining models (MM, MLP and CNN) have degraded performance for both augmentations.
In each of these cases, the RT-trained models suffer a smaller fractional reduction in their performance.
From the perspective of overall loss, the MLP and CNN models are each impacted by a similar amount, independent of augmentation strategy.
Indeed from this more global perspective, the U-trained models outperform the RT-trained models both in- and out-of-distribution data.
From this, we conclude that most of the variability reflected by the out-of-distribution dataset is not captured by our augmentations.

\begin{figure}[b!]
    \centering
    \includegraphics{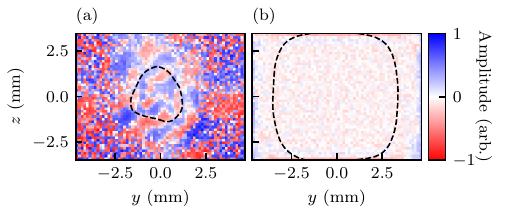}
    \caption{
    MM model kernels for temperature inference.
    The model is trained on (a) unaugmented data, and (b) data with combined reflection and translation augmentations.
    Kernels are for $\ex$ florescence images, and the black dashed curve encloses regions with significant florescence signal.
    }
    \label{fig:MNN_matrix}
\end{figure}

\section{Conclusion and outlook}
\label{sec:conclusion}

In this work, we explored the utility of machine learning (ML) techniques for extracting relevant characteristics of atoms in a magneto-optical trap (MOT), such as their number and temperature, from non-destructive fluorescence images.
We began by creating a labeled dataset with $\approx 39\times10^3$ elements from laser-cooled $\K39$ atoms in a MOT.
Each element of the dataset contains a pair of fluorescence images and a follow-up destructive absorption image acquired after a short time-of-flight.
Atom number and temperature labels were obtained from the absorption images.
We investigated five machine learning models, with a range of complexities, to estimate these parameters from the fluorescence images alone.
The training procedure optionally included data augmentation that combined reflections and translations of the fluorescence images.

Our first model, the trivial case with constant outputs (CON), served as the baseline to which the remaining models were compared.
The next two models were linear: a simple linear function (LIN) of the summed fluorescence counts, and a single fully connected layer (matrix multiplication, MM).
Both of these models improved upon the CON model for number inference, while only the MM model showed improvement for temperature inference.
The two non-linear models, a multi-layer perceptron (MLP) and a convolutional neural network (CNN), further improved inference for both number and temperature, with the CNN performing best and most robustly for all training and testing configurations.

\noindent{\it Data} --- The best CNN model predicts number with an uncertainty of $4\times10^6$ and temperature with a fractional uncertainty of $0.2$.
Interestingly, these regression uncertainties are below the estimated uncertainties of the labels, $6\times10^6$ and 0.5 respectively.
As noted in Sec.~\ref{sec:data_collection}a, the observed time-of-flight density distributions used for labeling are often poorly described by our simple Gaussian model, leading to inflated uncertainty estimates.
This confirms the opportunity for significant improvement in labeling.

\noindent{\it Training} --- During the final preparation of this manuscript, and after unblinding the test data, we initiated a $4\times10^5$ epoch training run for the fully augmented CNN model (10-fold cross-validation requires about 20 weeks on a single NVIDIA GeForce RTX 4080 16GB GDDR6X).
This shows that, while the loss first plateaus at about 300 epochs, both the training and validation loss begin dropping again after $10^4$ epochs.
After the full $4\times10^5$ epochs, the validation loss fell by an additional factor of five compared to the results presented above, and no sign of saturation was observed.
This underscores the potential improvements from model optimization and improved training methodology. 

\noindent{\it Explainability} --- The MM, MLP, and CNN models all utilized the atoms' spatial distribution for improved inference of both number and temperature.
Incorporating any form of translation augmentation into the training data degraded the MM model's capability for temperature inference to the level of the LIN and CON models, therefore implying that spatial structure is the only source of temperature information.

For each output, the MM model operates by learning a kernel that multiplies fluorescence images in a pixel-by-pixel manner.
Typical kernels for temperature inference are visualized in Fig.~\ref{fig:MNN_matrix}, both (a) without augmentation, and (b) with combined reflection and translation augmentation.
The dashed curves outline the region where fluorescence images show significant signal.
The kernel from unaugmented training data in Fig.~\ref{fig:MNN_matrix}(a) has significant spatial structure that is erased by the use of augmentation in (b), thereby recovering the simple summation employed in the LIN model.

Although these kernels directly visualize the MM model's operation, they do not suggest an underlying physical mechanism.
Furthermore, neither of the higher performing non-linear models can be interpreted even in this limited way.
This suggests the importance of exploring ML techniques targeting physical dynamics or interpretability, such as symbolic regression~\cite{Cranmer2023} or explainable boosting machines~\cite{Lou2013, Schug2025}, respectively.

\noindent{\it Outlook} --- Even on a low-end GPU, the models presented here can perform inference in $\approx0.5$~\si{\milli\second}.
This enables real-time applications, because it is far below the typical $\gtrsim 5$~\si{\milli\second} time-scale of MOT dynamics.
Such ML tools both provide diagnostic access to quantities that otherwise require destructive measurements, and open new pathways for real-time feedback control of laser-cooled atoms operating in novel parameter regimes~\cite{Gaudesius2021}.

The inherent complexity of quantum platforms---from system and state preparation, to control and finally measurement---makes them an ideal use case for ML-based information extraction and ML-enabled optimal control.
Our work is therefore a significant step in these directions, giving demonstrable access to otherwise hidden information, further motivating the use of ML methods in cold-atom-based platforms, and in quantum science and technology broadly speaking.

\subsection{Data Availability Statement}
The experimental datasets acquired and analyzed during the current study will be made publicly available upon publication.

\begin{acknowledgments}
The authors thank E.~B.~Norrgard and D.~Schug for carefully reading the manuscript.
F.~Salces-Carcoba created the CAD file used in Fig.~\ref{fig:workflow}a.
This work was partially supported by the National Institute of Standards and Technology and the National Science Foundation through the Physics Frontier Center at the Joint Quantum Institute (PHY-1430094) and the Quantum Leap Challenge Institute for Robust Quantum Simulation (OMA-2120757).
G.D.S. acknowledges the São Paulo Research Foundation (2024/20892-8).
\end{acknowledgments}

\bibliography{main}

\begin{thebibliography}{34}
\bibitem [{Equ()}]{Equipment}%
  \BibitemOpen
  \href@noop {} {}\bibinfo {note} {Any mention of equipment, instruments,
  software, or materials; it does not imply recommendation or endorsement by
  the National Institute of Standards and Technology.}\BibitemShut {Stop}%
\end{thebibliography}

\end{document}